\documentclass[prd,amsmath,amssymb,superscriptaddress,floatfix,nofootinbib,10pt]{revtex4}
\usepackage{times}
\usepackage{amssymb,amsbsy,amsmath,amsfonts}
\usepackage{graphicx}
\usepackage{float}
\usepackage{color}
\usepackage{morefloats}
\usepackage{rotating}
\usepackage{srcltx}
\usepackage{slashed}
\usepackage{multirow}
\usepackage{verbatim}
\usepackage{hyperref}
\usepackage{tabularx}
\usepackage{bm}
\allowdisplaybreaks[4]

\usepackage{bbding}
\usepackage{threeparttable}

\DeclareUnicodeCharacter{2212}{\ensuremath{-}}

\newcommand{\PreserveBackslash}[1]{\let\temp=\\#1\let\\=\temp}
\newcolumntype{C}[1]{>{\PreserveBackslash\centering}p{#1}}
\newcolumntype{R}[1]{>{\PreserveBackslash\raggedleft}p{#1}}
\newcolumntype{L}[1]{>{\PreserveBackslash\raggedright}p{#1}}

\begin{document}

\title{ Searching for the $2^+$ partner of the $T_{cs0}(2870)$ in the $B^- \to D^- D^0 K^0_S$ reaction
%The   $B^-\to D^-D^0K_S^0~(D^0\bar{K}^0)$    and the $I~J^P=0<~2^+$ state from $D^*\bar{K}^*$
}

\author{Jing Song}
\affiliation{School of Physics, Beihang University, Beijing, 102206, China}
\affiliation{Departamento de Física Teórica and IFIC, Centro Mixto Universidad de Valencia-CSIC Institutos de Investigación de Paterna, 46071 Valencia, Spain}

\author{Zi-Ying Yang}
\affiliation{School of Physics, Beihang University, Beijing, 102206, China}
\affiliation{Departamento de Física Teórica and IFIC, Centro Mixto Universidad de Valencia-CSIC Institutos de Investigación de Paterna, 46071 Valencia, Spain}

\author{ Eulogio Oset}
\email[]{oset@ific.uv.es}
\affiliation{Departamento de Física Teórica and IFIC, Centro Mixto Universidad de Valencia-CSIC Institutos de Investigación de Paterna, 46071 Valencia, Spain}
\affiliation{Department of Physics, Guangxi Normal University, Guilin 541004, China}

\begin{abstract}
 We study the $B^- \to D^- D^0 K^0_S$ reaction, recently analyzed by the LHCb collaboration, where a clear signal for the exotic $T_{cs0}(2870)$  state has been reported. We call the attention to a small peak in the $D^0 K^0_S$ mass distribution that could correspond to a state of the same nature as the $T_{cs0}(2870)$ ($D^* \bar K^*$ nature in the molecular picture) but with $J^P= 2^+$. In order to magnify the signal for the state, we calculate the moments of the angle-mass distribution, which are linear in the resonance signal, rather than quadratic for the angle integrated mass distribution. We find spectra for the moments with a strength far bigger than that for the angle integrated mass distribution, which should encourage the evaluation of these moments from the present measurements of the reaction. 

\end{abstract}

%\pacs{13.75.Ev,12.39.Fe,21.30.Fe}
%\keywords{}

\maketitle

\section{Introduction}
 The $B^- \to D^- D^0 K^0_S$ reaction has been recently measured by the LHCb collaboration~\cite{LHCb:2024xyx} and in the $D^0 K^0_S$ mass distribution a clean peak is seen around 2900 MeV, which is identified as the $T_{cs0}(2870)$.
 This state, of $0^+$ nature, originally called $X_0(2900)$, was first reported by the LHCb collaboration  in the $B^+ \to D^+ D^−K^+$ decay \cite{LHCb:2020bls,LHCb:2020pxc}, together with the $X_1(2900)$ state of $1^-$ nature, by looking at the $D^−K^+$ mass distribution. 
 These meson states, containing $\bar{c}\bar{s}$ quarks, are manifestly exotic since they cannot be accommodated with the standard $q \bar{q}$ picture of mesons. Actually, the state  $T_{cs0}(2870)$ was predicted with a good approximation for the mass and width in~\cite{Molina:2010tx}, as a molecular state of $\bar{D}^*K^*$ nature. 
 An update of this work to the light of the LHCb data is done in~\cite{Molina:2020hde}.  Since then, many theoretical works have been produced trying to explain the nature of the $T_{cs0}(2870)$ state as molecular state~\cite{LHCb:2022lzp,Karliner:2020vsi,He:2020jna,Guo:2021mja,Ozdem:2022ydv,Wang:2020xyc,Agaev:2021knl,Chen:2020aos,He:2020btl,Burns:2020epm,  Kong:2021ohg,Ke:2022ocs,Yu:2023avh,Ding:2024dif}, or as a compact tentraquak state \cite{Agaev:2021knl,Yang:2021izl,Wang:2020prk,Hu:2020mxp,Yang:2020atz,Tan:2020cpu,Agaev:2022eeh,Wei:2022wtr,Liu:2022hbk,Ortega:2023azl}.   Different reactions have also been proposed to determine the nature of the states \cite{Chen:2020eyu,Burns:2020xne,Abreu:2020ony,Bayar:2022wbx,Lin:2022eau, Yang:2024coj}. A recent updated review of the new hadron states can be seen in~\cite{Chen:2022asf}. Further discussions concerning the $T_{cs0}(2870)$ and related states can be seen in~\cite{Bayar:2022wbx}.
 
\begin{figure}[H]
  \centering
\includegraphics[width=0.35\textwidth]{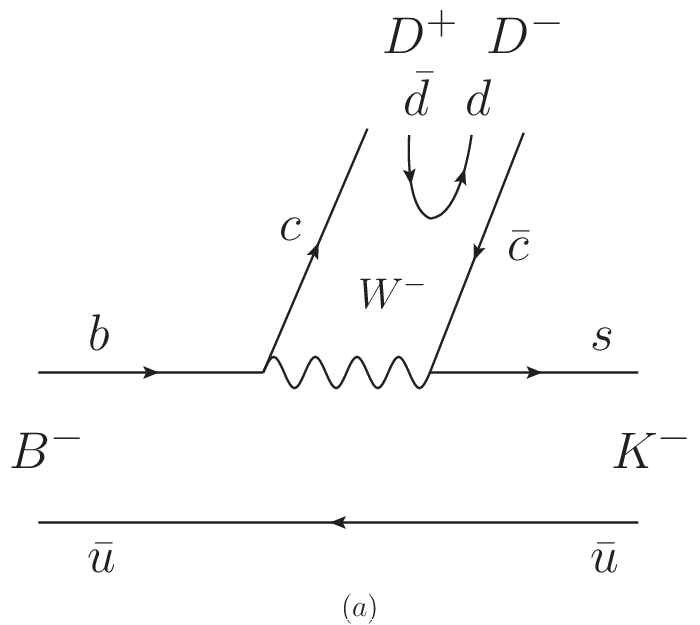}
\includegraphics[width=0.35\textwidth]{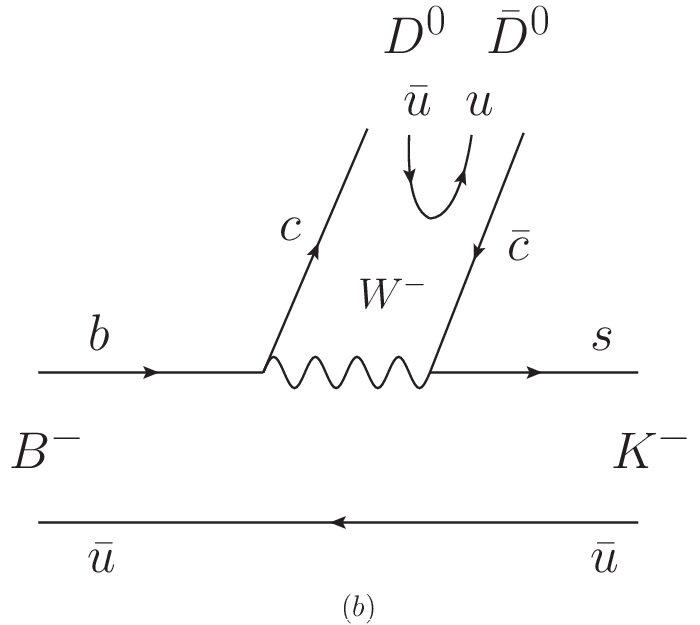}
  \caption{Mechanisms for $B^- \to D^- D^+K^-$ (a), $B^- \to D^0\bar{D}^0K^-$ (b).}\label{fig1}
\end{figure}

\begin{figure}[H]
  \centering
\includegraphics[width=0.35\textwidth]{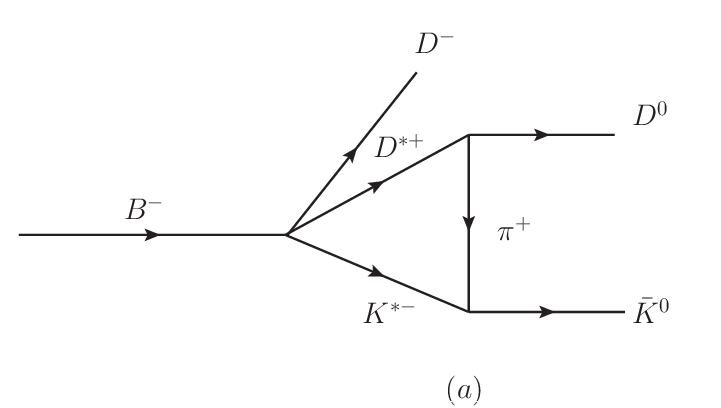}
\includegraphics[width=0.55\textwidth]{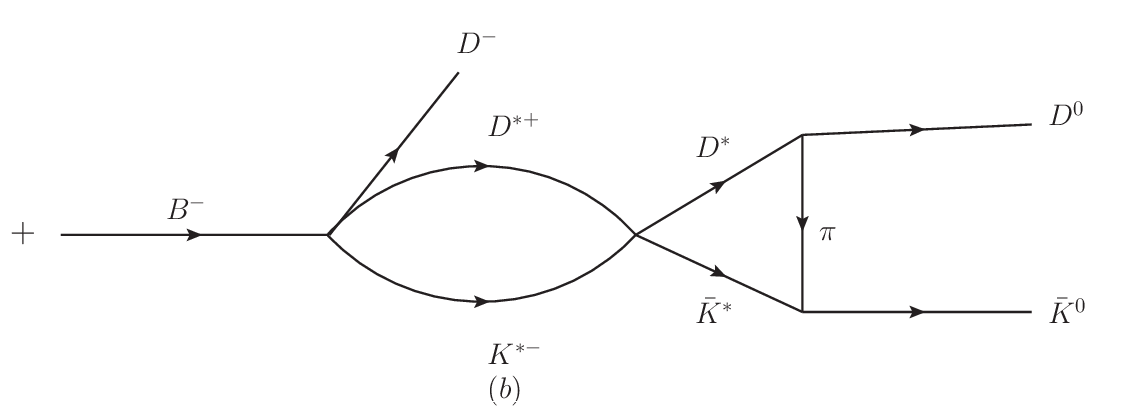}
  \caption{ $B^- \to D^-D^{*+}K^{*-} \to D^-D^0\bar{K}^0$ through reactions. (a) direct $D^{*+}K^{*-} \to D^0\bar{K}^0$ transition;  (b) resonant $D^{*+}K^{*-} \to D^0 \bar K^0$ transition mediated by the $T_{cs0}(2870)$.}\label{fig2}
\end{figure}

 Coming back to the recent reaction $B^- \to D^- D^0 K^0_S$, which is actually $B^- \to D^- D^0 \bar K^0$, with the $\bar K^0$ observed as a $K^0_S$, one might think that this reaction is basically the same as the 
 $B^+ \to D^+ D^−K^+$ (or $B^- \to D^- D^+K^-$), simply substituting $D^+K^-$ by $D^0 \bar K^0$, which are members of the same isospin zero multiple. Then one might expect similar spectra in the mass distributions, which is actually not the case. In fact, in the $B^+ \to D^+ D^−K^+$, the LHCb collaboration reported two states, the 
$X_0(2900)$ and the $X_1(2900)$. However, in the $B^- \to D^- D^0 K^0_S$ reaction, only a signal of the $X_0(2900)$ state is seen. We can qualitatively  understand this based on the mechanisms for the decay, which are actually quite different. 
Indeed, the $B^- \to D^- D^+K^-$ reaction proceeds naturally via internal emission~\cite{Chau:1982da}, as seen in Fig.~\ref{fig1} (a), with hadronization 
\noindent of the $c\bar{c}$ component into $D^+D^-$. Similarly, we can produce $D^0\bar{D}^0$, but then one has the final state $D^0\bar{D}^0K^-$, not the desired $D^-D^0\bar{K}^0$ state. We need final state interaction to make the transition $\bar{D}^0K^-\to D^-\bar{K}^0$ and have finally the $D^-D^0\bar{K}^0$. 
We could equally start from $  D^+D^-K^-$ with   the transition $D^+K^- \to D^0\bar{K}^0$, but in both cases we need the final state interaction to produce the desired final state.
This makes the reactions different from the very beginning. From our perspective, on top of some background, we would also be producing for instance $D^{*+}D^-K^{*-}$ and then we would have to make a transition to $D^0D^-\bar{K}^0$. This can be done with the reaction mechanism depicted in Fig.~\ref{fig2}. We can have
\noindent direct transition of $D^{*+}K^{*-}\to D^0\bar{K}^0$ as shown in Fig.~\ref{fig2} (a), but once the $D^{*+} K^{*-}$ is produced, its interaction is inevitable and the $D^{*}\bar K^{*}$ interaction in $J=0$ and $S-$wave, which is allowed by the weak production, produces the $X_0(2900)$ ($T_{cs0}(2870)$). The fact that the $D^-D^0\bar{K}^0$ state requires rescattering to be produced, certainly favors the production of the $T_{cs0}(2870)$ in the reaction, which could explain why the $T_{cs0}(2870)$ plays a more important role in the   
 $B^- \to D^-D^0\bar{K}^0$  reaction than in $B^- \to D^-D^+K^-$.

 This said, the purpose of the present work is to investigate if from the data of   $B^- \to D^-D^0\bar{K}^0$ one could induce the existence of a $2^+$ state of $D^{*}\bar K^{*}$  nature. The $D^{*}\bar K^{*}$  can lead to three states in $S-$wave. Indeed, three bound states were predicted in~\cite{Molina:2010tx} which have been corroborated by a large number of theoretical works mentioned above studying the $T_{cs0}(2870)$ as a molecular state of $D^{*}\bar K^{*}$.  The $1^+$ state cannot decay to   $D^{0}\bar K^{0}$ for reasons of spin and parity conservation. Thus, only the $2^+$ state could show up in the reaction. However, the transition of $D^{*}\bar K^{*}$ with $J=2$ to   $D^{0}\bar K^{0}$ requires $L=2$ in   $D^{0}\bar K^{0}$, and one can expect a strong reduction in the signal of this state. In such a case the evaluation of the moments $\frac{d \Gamma_l}{d M_{\mathrm{inv}}} = \int d \tilde{\Omega} \frac{d \Gamma}{d M_{\mathrm{inv}} d \tilde{\Omega}} Y_{l m}$ is very useful, since some of these moments are linear in the production amplitude of the resonance, while in ${d \Gamma}/{d M_{\mathrm{inv}}}$ the signal appear quadratic and small. This was the case in the work of \cite{Bayar:2022wbx} where a small signal in $d\Gamma_0/dM_\text{inv}(D^- {K}^+)$  was identified around $2775$~MeV, where the mass of the $2^+$ was predicted in~\cite{Molina:2020hde}, which  gave rise to a clear signal in   $d\Gamma_3/dM_\text{inv}(D^- {K}^+)$ through interference with the $X_1(2900)$ signal of the experiment. The calculated moment actually looks remarkably similar to the one observed in~\cite{LHCb:2020pxc} (compare Fig.4 of \cite{Bayar:2022wbx} with Fig.19 of \cite{LHCb:2020pxc}).
 
 The idea here is the same as in~\cite{Bayar:2022wbx}, and we identify a small peak (which actually is compatible with a statistical fluctuation), assuming that this reflects a small signal for the $2^+$ state and then we shall calculate the moments of the distribution. The method of the moments was also used in~\cite{Lyu:2024zdo}  to get a signal of the $I=1, J^P=2^+$ state of $D^*K^*\to D_S^*\rho$ nature in the $B^+\to D^{*-}D^+K^+$ reaction, which is predicted in~\cite{Molina:2010tx} (see \cite{Molina:2022jcd} for references on this issue). It has also been used in~\cite{Song:2024dan} to observe the same state in the $\Lambda_b\to \Sigma_c^{++}D^-K^-$ reaction,
and more recently in the  $B^+\to D^{*-}D_s^+\pi^+$   reaction, looking again for this $2^+$ state~\cite{Lyu:2025rsq}.
 
 While the idea used here is the same as in~\cite{Bayar:2022wbx}, there are again large difference, because the interference in the moments is now between the $J=0$ and the $J=2$ signal, while in the   $B^- \to D^-D^+K^-$  reaction the interference was between the $J=2$
and  $J=1$  signals. Thus, the moments where interference is expected are different. We shall also calculate the moments assuming that the small signal is due  to a state of $J=0$ or $J=1$, which lead to different structures from  those of $J=2$. The reconstruction of the moments from the data of~\cite{LHCb:2024xyx} should provide an answer to the question of which spin has to be attributed to the small peak.

\section{formalism}
Following the formalism of~\cite{Bayar:2022wbx}
 we write the transition amplitude for  $B^- \to D^-D^{0}\bar K^{0}$ as 
 $$
 t=aY_{00}(\theta)+bY_{20}(\theta)+cY_{10}(\theta)
 $$
where $\theta$ is the angle of $\bar K^0$  with the $D^-$ in the rest frame of  $D^{0}\bar K^{0}$. 
The $a$ term will account for the $T_{cs0}(2870)$ resonance, and has another part accounting for the case that  the small peak corresponds to a $J=0$ resonance. The $b$ term would account for the case where that peak corresponds to a state of $J=2$
and the $c$ term for the case that it would be a state of $J=1$. We should note that the case of $J=1$ cannot come from decay of a $D^{*}K^{*}$  state, but could come from other sources. 

Then we write 
\begin{align}\label{6_1}
 &a = a_0 +a_0'~\frac{M_B^2}{M_\text{inv}^2(D^0\bar K^0)-M_{R_0}^2+iM_{R_0}\Gamma_{R_0}} +a_0''~\frac{M_B^2}{M_\text{inv}^2(D^0\bar K^0)-M_{X_0}^2+iM_{X_0}\Gamma_{X_0}},\nonumber\\
 &b = b'~\frac{\tilde{k}^2}{M_\text{inv}^2(D^0\bar K^0)-M_{R_2}^2+iM_{R_2}\Gamma_{R_2}},\nonumber\\
 &c = c'~\frac{M_B\tilde{k}}{M_\text{inv}^2(D^0\bar K^0)-M_{R_1}^2+iM_{R_1}\Gamma_{R_1}}. 
\end{align}
The factor $M_B^2$ in the term $a$ is added to have $a_0',~a_0''$ of the same dimensions of $a_0$ and the terms with $\tilde{k},~\tilde{k}^2$ reflect the $P-$ and $D-$ waves in the cases of $J=1,~J=2$. The magnitude $\tilde{k}$ is the $\bar K^0$ momentum in the $D^0\bar K^0$ rest frame, given by 
\begin{align}
& \tilde{k} = \frac{\lambda^{1 / 2}\left(M_{\text{inv}}^2(D^0\bar K^0), m_{D^0}^2, m_{\bar K^0}^2\right)}{2 M_{\text{inv}}(D^0 \bar K^0)}.
\end{align}
The moments are defined as 
\begin{align}
\frac{d \Gamma_l}{d M_{\mathrm{inv}}(D^0\bar K^0)} = \int d \tilde{\Omega} \frac{d \Gamma}{d M_{\mathrm{inv}}(D^0\bar K^0) d \tilde{\Omega}} Y_{l 0},
\end{align}
where 
\begin{align}\label{6_2}
    \frac{d \Gamma}{d M_{\mathrm{inv}}(D^0\bar K^0) d \tilde{\Omega}}=\frac{1}{(2\pi)^4}\frac{1}{8M_B^2}p_{D^-}\tilde{k}\sum|t|^2,
\end{align}
with 
\begin{align}
p_{D^-} = \frac{\lambda^{1 / 2}\left(M_B^2, M_{D^-}^2, M_{\text{inv}}^2(D^0\bar K^0)\right)}{2 M_B},
\end{align}
and $\tilde{\Omega}$  is the solid angle of $\bar K^0$ referred to the $D^-$  momentum in the $D^0\bar K^0$ rest frame. $|t|^2$ in Eq.~(\ref{6_2}) is given by 
\begin{align}
    |t|^2= |a|^2Y_{00}^2+|b|^2Y_{20}^2+|c|^2Y_{10}^2+2 \text{Re}(ab^*)Y_{00}Y_{20}+2 \text{Re}(ac^*)Y_{00}Y_{10}+2 \text{Re}(bc^*)Y_{20}Y_{10}.
\end{align}
Then, using the standard formulas for integrals of three spherical harmonics one immediately obtains ($M_{\mathrm{inv}} \equiv M_{\mathrm{inv}}(D^0\bar K^0)$),
\begin{align}\label{7_1}
&\frac{d\Gamma_0}{dM_{\text{inv}}} = FAC \left[ |a|^2 + |b|^2 + |c|^2 \right],\nonumber\\
&\frac{d\Gamma_1}{dM_{\text{inv}}} = FAC \left[ 2 \operatorname{Re}(ac^*) + \frac{2}{\sqrt{5}} 2 \operatorname{Re}(bc^*) \right],\nonumber\\
&\frac{d\Gamma_2}{dM_{\text{inv}}} = FAC \left[ \frac{2}{7} \sqrt{5} |b|^2 + \frac{2}{5} \sqrt{5} |c|^2 + 2 \operatorname{Re}(ab^*) \right],\nonumber\\
&\frac{d\Gamma_3}{dM_{\text{inv}}} = FAC \sqrt{\frac{15}{7}} \frac{3}{5} 2 \operatorname{Re}(bc^*),\nonumber\\
&\frac{d\Gamma_4}{dM_{\text{inv}}} = FAC \frac{6}{7} |b|^2,
\end{align}
where
\begin{align}
FAC = \frac{1}{\sqrt{4 \pi}} \frac{1}{(2\pi)^4}\frac{1}{8M_B^2}p_{D^-}\tilde{k},
\end{align}
and then 
\begin{align}
    \frac{d\Gamma}{dM_{\text{inv}}} = \sqrt{4 \pi} \frac{d\Gamma_0}{dM_{\text{inv}}},
\end{align}
which agree with the formulas found in \cite{LHCb:2016lxy} up to a global factor (see \cite{Bayar:2022wbx}).

\section{results}
We evaluate the mass distributions and moments assuming three different scenarios.
In all cases,we take $M_{X_0},~\Gamma_{X_0}$ from the PDG~\cite{ParticleDataGroup:2024cfk} as $M_{X_0}=2866$~MeV, $\Gamma_{X_0}=57$~MeV.

\textbf{Case 1:} J=0 case.

We assume that the small peak is due to a state with $J=0$. In this case we have in Eq.~(\ref{6_1}), $b'=c'=0$. Then we adjust $a_0',~a_0'',~M_{R_0},~\Gamma_{R_0}$, to $\frac{d\Gamma}{dM_{\text{inv}}(D^0\bar K^0)}$  (which we take as the counts in the experiment), with $M_{R_0},~\Gamma_{R_0}$ standing the mass and width of the $R_0$ resonance with $J=0$.

\textbf{Case 2:} J=1 case.

We assume that the small peak corresponds to $J=1$. In this case we have in Eq.~(\ref{6_1}), $a_0'=0, ~b'=0$, and do a fit to $\frac{d\Gamma}{dM_{\text{inv}}(D^0\bar K^0)}$  to determine  $a_0'',~c',~M_{R_1},~\Gamma_{R_1}$,  the last two parameters standing for the mass and width of the $J=1,~R_1$ resonance.  

\textbf{Case 3:} J=2 case.

Here we assume that the small peak corresponds to $J=2$, our favorite choice. Here we take $a_0'=0, ~c'=0$ and adjust $\frac{d\Gamma}{dM_{\text{inv}}(D^0\bar K^0)}$   to the data to determine  $a_0'',~b',~M_{R_2},~\Gamma_{R_2}$.

In all cases, we have taken the parameter $a_0$ as a background in $S-$wave to provide the curve for $\frac{d\Gamma}{dM_{\text{inv}}(D^0\bar K^0)}$ provided in  Fig.2 (b) of the experiment in \cite{LHCb:2024xyx} in the absence of the $T_{cs0}(2870)$ contribution. As we can see in that figure, this background is basically given by a $S-$wave contribution. The small $P-$wave contribution for Fig.2 (a) of \cite{LHCb:2024xyx} for the $D^-K_S^0$  mass distribution from background and the $D_{s1}(2700)$ resonance gets diluted into other partial waves when making a boost to the $D^0\bar K^0$  rest frame. We find a parabola to be sufficient good for our purposes as
\begin{align}
  \sqrt{4 \pi}~  FAC~a_0^2 = \alpha+\frac{\beta}{m_K^2}(M_{\text{inv}}-\sqrt{s}_0)^2
\end{align}
 and we choose $\sqrt{s}_0=2870$~MeV to agree with the background of~\cite{LHCb:2024xyx}, and  $ \alpha,~ \beta$ are also let free in the fits to the data.

\begin{figure}[H]
  \centering
\includegraphics[width=0.45\textwidth]{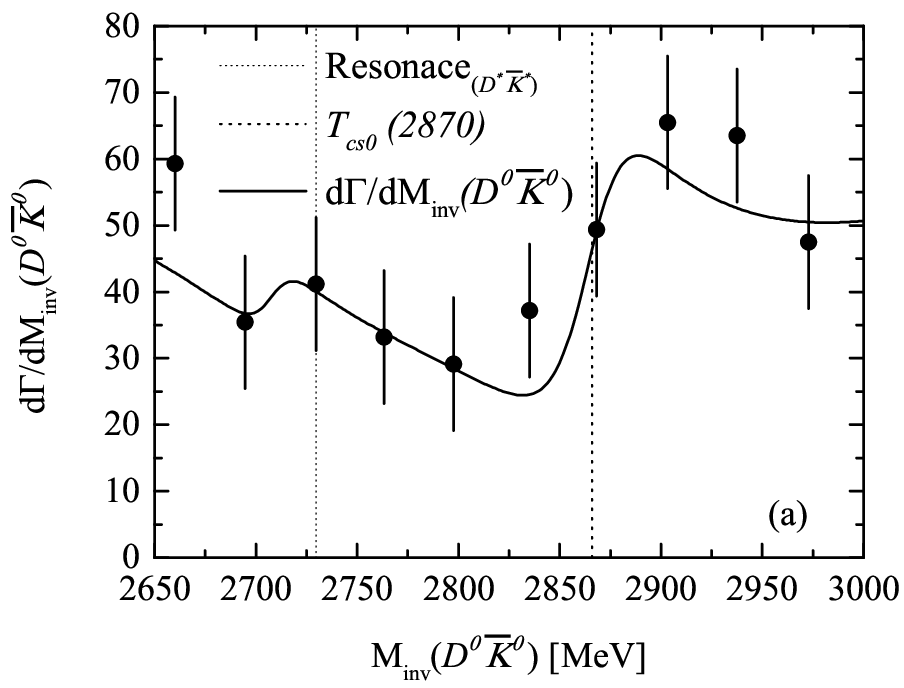}
\includegraphics[width=0.45\textwidth]{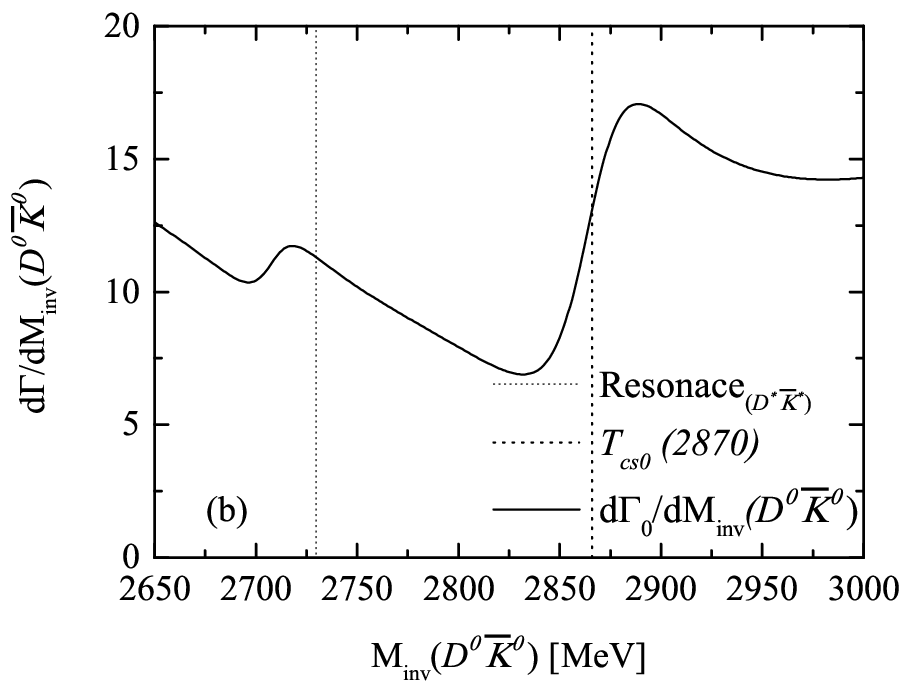}
  \caption{\textbf{Case 1} ($J=0$): The mass distributions of  $d\Gamma/dM_\text{inv}(D^0\bar{K}^0)$ (a) and $d\Gamma_0/dM_\text{inv}(D^0\bar{K}^0)$ (b) with fitted parameters.}\label{fig3}
\end{figure}

 In Fig.~\ref{fig3} (a) we plot   $\frac{d\Gamma}{dM_{\text{inv}}(D^0\bar K^0)}$   for case 1 ($J=0$) with the experimental points in the region of interest. There one sees the  $T_{cs0}(2870)$ and the possible new state at lower energy. The fit to the data is fair and we show the parameters obtained in Table~\ref{table1}.  The mass obtained for the new resonance is around $2708$~MeV and the width around $32$~MeV. While the width is in line with the results obtained in~\cite{Molina:2020hde} for the $2^+$ state, the mass is low compared with the $2778$~MeV prediction. In Fig.~\ref{fig3} (b) we show  $\frac{d\Gamma_0}{dM_{\text{inv}}(D^0\bar K^0)}$   which is actually the same curve as Fig.~\ref{fig3} (a) divided by $\sqrt{4\pi}$. Hence this gives no extra information.
The background obtained is qualitatively similar to the one found in~\cite{LHCb:2024xyx}, and so is it in the other cases.

\begin{figure}[H]
  \centering
\includegraphics[width=0.45\textwidth]{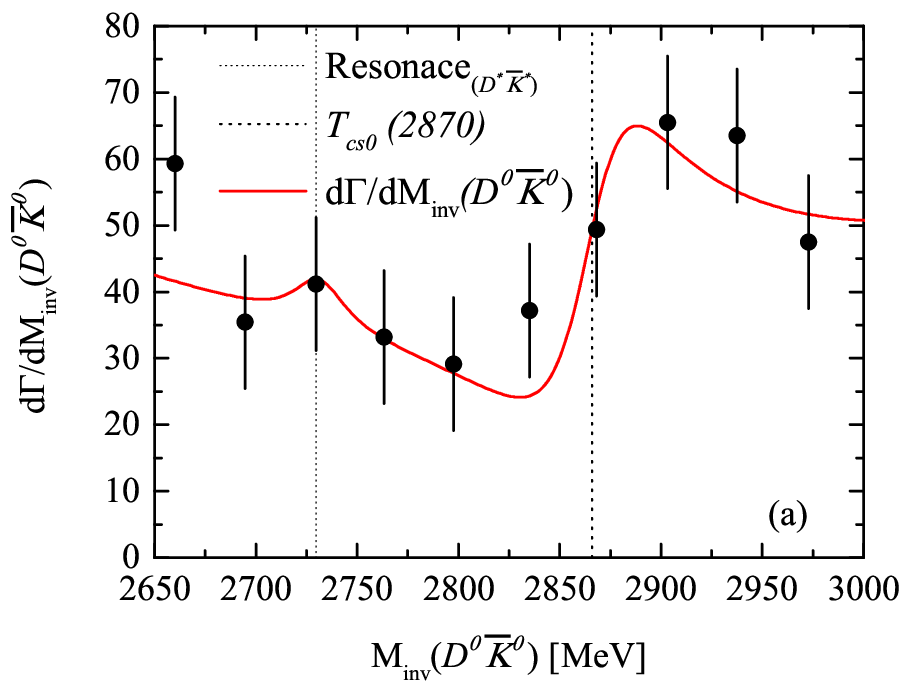}
\includegraphics[width=0.45\textwidth]{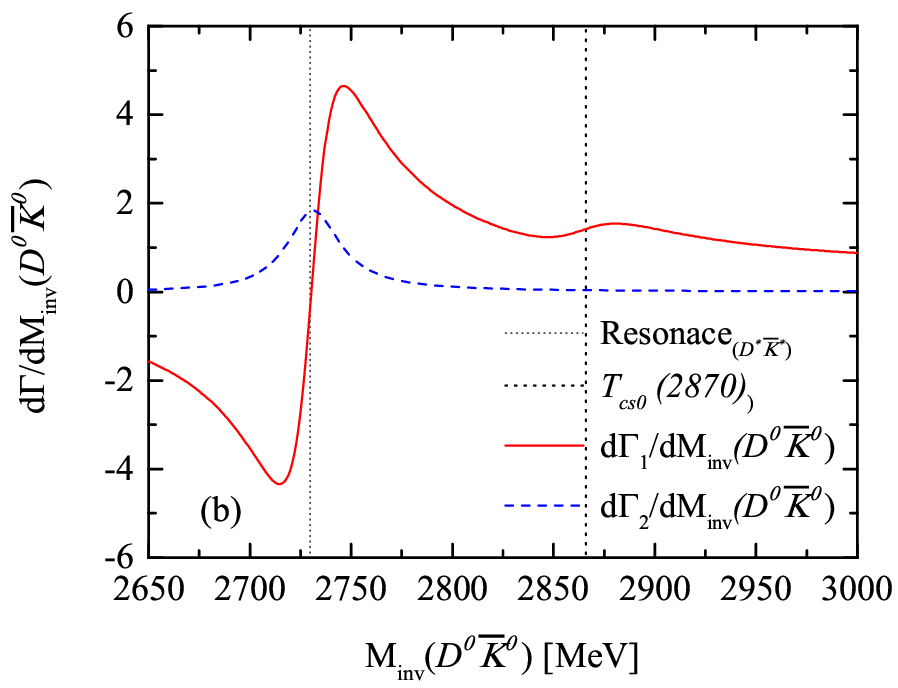}
  \caption{\textbf{Case 2} ($J=1$): The mass distributions of  $d\Gamma/dM_\text{inv}(D^0\bar{K}^0)$ (a) and $d\Gamma_1/dM_\text{inv}(D^0\bar{K}^0)$,~$d\Gamma_2/dM_\text{inv}(D^0\bar{K}^0)$ (b) with fitted parameters.}\label{fig4}
\end{figure}

 In Fig.~\ref{fig4} (a) we show the results for case 2 ($J=1$). We also get a very similar fit to $\frac{d\Gamma}{dM_{\text{inv}}(D^0\bar K^0)}$ and in Table~\ref{table2} we show the parameters of the fit. The   width of the new state is basically the same as in the former case, but the mass is now $2731$~MeV, closer to the one found in~\cite{Molina:2020hde}. Furthermore, in Fig.~\ref{fig4} (b) we show the moments  $\frac{d\Gamma_1}{dM_{\text{inv}}(D^0\bar K^0)}$ and  $\frac{d\Gamma_2}{dM_{\text{inv}}(D^0\bar K^0)}$. We see that  ${d\Gamma_2}$ shows only a small peak around the new resonance, since the $T_{cs0}(2870)$    signal is killed in this moment. However, we observe that $\frac{d\Gamma_1}{dM_{\text{inv}}(D^0\bar K^0)}$ shows an interference structure around the peak of the new resonance, with a strength far larger than the contributions to the mass distribution (the term $2 FAC|c|^2/\sqrt{5}$  of Eq.~(\ref{7_1})). Normalizing the $|c|^2$ term to give    $\frac{d\Gamma}{dM_{\text{inv}}(D^0\bar K^0)}$ of  Fig.~\ref{fig4} (a), the signals get multiplied by the factor $\sqrt{5\pi}\approx 4$, indicating that the signal for $\frac{d\Gamma_1}{dM_{\text{inv}}(D^0\bar K^0)}$ is very large and  should be perfectly visible. 

\begin{figure}[H]
  \centering
\includegraphics[width=0.45\textwidth]{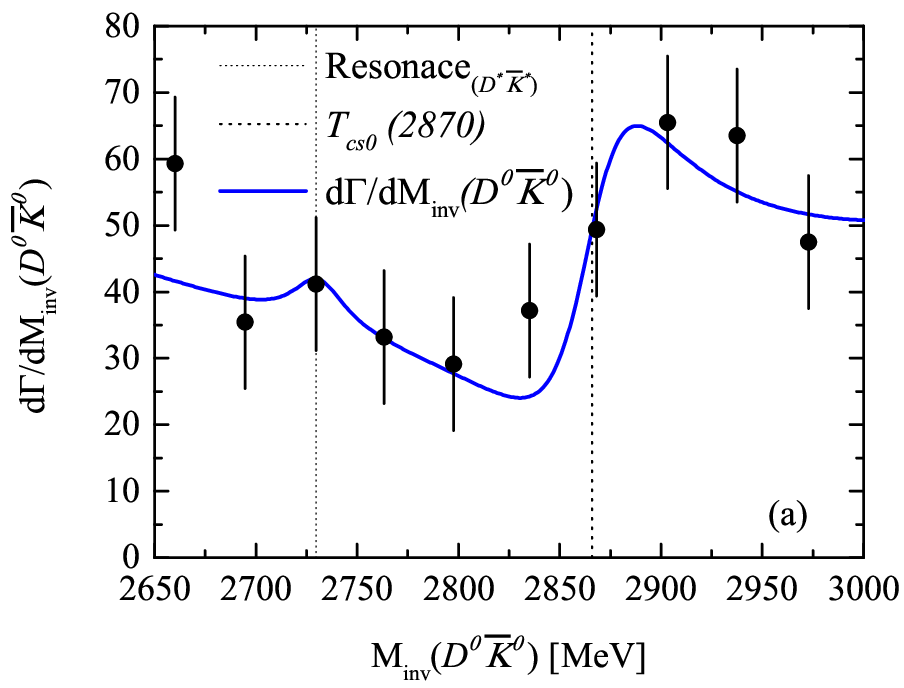}
\includegraphics[width=0.45\textwidth]{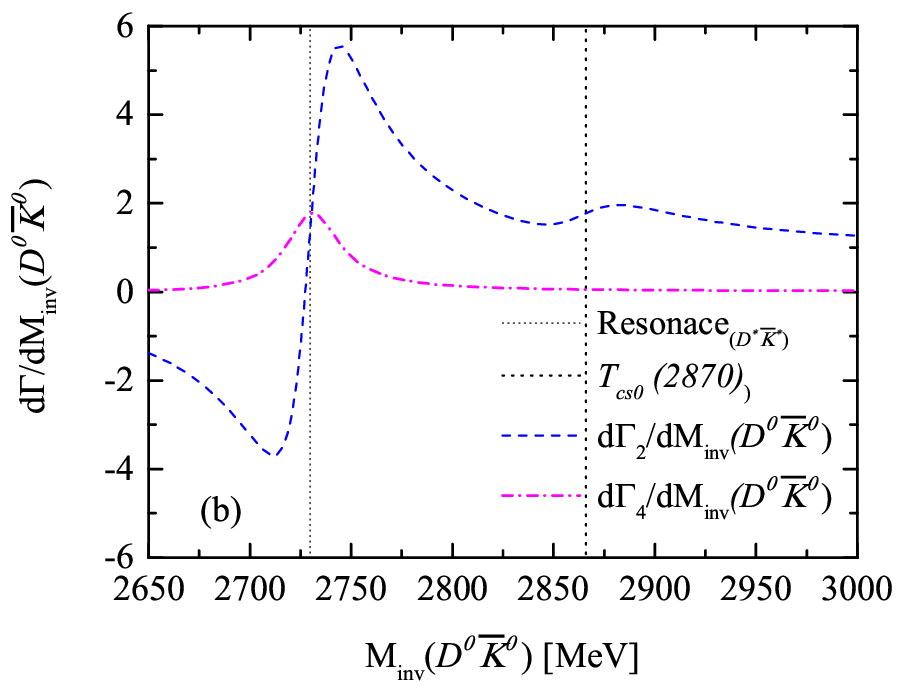}
  \caption{\textbf{Case 3} ($J=2$): The mass distributions of  $d\Gamma/dM_\text{inv}(D^0\bar{K}^0)$ (a) and $d\Gamma_2/dM_\text{inv}(D^0\bar{K}^0)$,~$d\Gamma_4/dM_\text{inv}(D^0\bar{K}^0)$ (b) with fitted parameters.}\label{fig5}
\end{figure}

 In Fig.~\ref{fig5} (a) we show again $\frac{d\Gamma}{dM_{\text{inv}}(D^0\bar K^0)}$ for case 3 ($J=2$), which is  similar to the other cases, and in  Fig.~\ref{fig5} (b) we show the moments $\frac{d\Gamma_2}{dM_{\text{inv}}(D^0\bar K^0)}$ and $\frac{d\Gamma_4}{dM_{\text{inv}}(D^0\bar K^0)}$, with  parameters  listed in Table~\ref{table3}. Once again the signal for $d\Gamma_2$, which contains the interference of the $T_{cs0}(2870)$  and the new resonance, gets magnified, which again indicates that the signal should be clearly visible.

 We observe that, depending on the assumed spin for the new resonance, the moments which show the strong interference are different. The evaluation of these moments from the experimental mass distributions would then help to determine the spin of the new resonance. Certainly, a better statistics for the experiments, which should be available in future runs of the LHCb, would greatly help to find out about this elusive $2^+$ state which is obtained in many theoretical papers.

\begin{table}[H]
%\footnotesize
\centering
\caption{ After best fit: The value of parameters $\alpha$, $\beta$, $a_0'$, $a_0''$, $M_{R0}$ and $\Gamma_{R0}$ for \textbf{Case 1} ($J=0$)  [units of $M_{R0}$ and $\Gamma_{R0}$ in MeV]. }
\label{table1}
\setlength{\tabcolsep}{32pt}
\begin{tabular}{ccccccc}
\hline \hline
$\alpha$ & $\beta$   & $a_0'$ & $a_0''$ & $M_{R0}$ & $\Gamma_{R0}$ \\
\hline 
$37$    & $77$  &   $1$  &  
$ 11  $ & $  2708 $	 & $  32 $ \\
\hline \hline
\end{tabular}
\end{table}

\begin{table}[H]
%\footnotesize
\centering
\caption{ After best fit: The value of parameters $\alpha$, $\beta$, $a_0''$, $c'$, $M_{R1}$ and $\Gamma_{R1}$ [units  of $M_{R1}$ and $\Gamma_{R1}$ in MeV] for \textbf{Case 2} ($J=1$). }
\label{table2}
\setlength{\tabcolsep}{32pt}
\begin{tabular}{ccccccc}
\hline \hline
$\alpha$ & $\beta$   & $a_0''$ & $c'$ & $M_{R1}$ & $\Gamma_{R1}$ \\
\hline 
 $39$  & $44$  & $12$    &  
$ 50 $ & $   2731$	 & $  32  $ \\
\hline \hline
\end{tabular}
\end{table}

\begin{table}[H]
%\footnotesize
\centering
\caption{ After best fit: The value of parameters $\alpha$, $\beta$, $a_0''$,  $b'$, $M_{R2}$ and $\Gamma_{R2}$ [units  of $M_{R2}$ and $\Gamma_{R2}$ in MeV] for \textbf{Case 3} ($J=2$). }
\label{table3}
\setlength{\tabcolsep}{32pt}
\begin{tabular}{ccccccc}
\hline \hline
$\alpha$ & $\beta$ & $a_0''$ & $b'$ & $M_{R2}$ & $\Gamma_{R2}$ \\
\hline 
$39$  &   $45$   & $12$   & 
$461 $ & $   2730  $	 & $   32 $ \\
\hline \hline
\end{tabular}
\end{table}

\section {Conclusions}
We have looked into the $B^- \to D^- D^0 K^0_S$ reaction comparing it to   $B^- \to D^- D^+K^-$. We have made a qualitative discussion trying to understand why in the latter reaction the two exotic states $X_0(2900)$ ($T_{cs0}(2870)$) and the $X_1(2900)$ ($T_{cs1}(2900)$) were observed, while in the former one only the $T_{cs0}(2870)$ is observed. We showed that in spite of the obvious similarity of the two reactions, since $D^+K^-$  and $D^0 \bar K^0$ belong to the same isospin multiplet, the two reactions proceed differently in the weak decay, where $B^- \to D^- D^+K^-$proceeds directly through internal emission, while $B^- \to D^- D^0 K^0_S$ requires necessarily final state interaction after the first weak decay products. One can then produce  $B^-\to D^-D^{*+}K^{*-}$ and the $D^{*+}K^{*-}$, after rescattering in $S-$wave producing the $T_{cs0}(2870)$, can make a transition to 
 $D^0 \bar K^0$ to produce the desired final state. 
    
    After this realization, the paper concentrates on the possibility of extracting information from the $D^0 K^0_S$ mass distribution of the $B^- \to D^- D^0 K^0_S$ reaction on the existence of a $2^+$ state of $D^* \bar K^*$ nature. This state is predicted in many theoretical works, together with the $0^+$, which corresponds to $T_{cs0}(2870)$, and another state of $1^+$ which cannot decay to $D \bar K$. Claims of evidence for this $2^+$ state were already made from the analysis of the $B^- \to D^- D^+K^-$ reaction by looking at the  $D^+K^-$ mass distribution and the momentum $d\Gamma_3/ d M_\text{inv}$ of this distribution \cite{Bayar:2022wbx}. 
    Here we have followed the same steps as in this latter work and also find a likely signal for this state in a small peak of the $D^0 K^0_S$ mass distribution in the $B^- \to D^- D^0 K^0_S$ reaction. After that, we evaluate the moments of the angular-mass distribution and make predictions assuming that this peak could correspond to a state of $J= 0, 1, 2$. Once again we show that these moments are quite different from those predicted~\cite{Bayar:2022wbx}  and observed~\cite{LHCb:2020pxc} in the $B^- \to D^- D^+K^-$ reaction.   We stress that the moments, which are linear in the small signal of the new state, show a spectrum far clearer than the peak observed in the angle integrated mass distribution, which is quadratic in the small resonance signal. In particular, should the state correspond to a $2^+$ state, we predict a clean spectrum for the $d \Gamma_2/d M_\text{inv}$ moment. We make a call for the construction of the moments for the $B^- \to D^- D^0 K^0_S$ reaction, as was already done for  $B^- \to D^- D^+K^-$, with the hope that the signals predicted can be verified and evidence for some new state is observed. Certainly, a better statistics for the reaction, which should be available in future runs of the LHCb, would greatly help clarifying this situation.   
Actually, with the present statistics and separation of experimental points, one cannot exclude that the small peak  discussed here could be a statistical fluctuation, or that some other small peak could not have been observed. However, the work carried out here, together with the analysis of~\cite{Bayar:2022wbx} for the $B^- \to D^- D^+ K^-$    reaction  and the coincidence of the $d\Gamma_3$ moments in~\cite{Bayar:2022wbx} and the LHCb experiment~\cite{LHCb:2020bls}, should
provide sufficient motivation to study the moments of the $B^- \to D^- D^0 K^0_S$ reaction and also come back to it when more statistics is available in the future.

\section{Acknowledgments}
This work is partly supported by the National Natural Science
Foundation of China under Grants  No. 12405089 and No. 12247108 and
the China Postdoctoral Science Foundation under Grant
No. 2022M720360 and No. 2022M720359. ZY.Yang and J. Song wish to thank support from the China Scholarship Council. This work is also supported by
the Spanish Ministerio de Economia y Competitivi-
dad (MINECO) and European FEDER funds under
Contracts No. FIS2017-84038-C2-1-P B, PID2020-
112777GB-I00, and by Generalitat Valenciana under con-
tract PROMETEO/2020/023. This project has received
funding from the European Union Horizon 2020 research
and innovation programme under the program H2020-
INFRAIA-2018-1, grant agreement No. 824093 of the
STRONG-2020 project. This work is supported by the Spanish Ministerio de Ciencia e Innovaci\'on (MICINN) under contracts PID2020-112777GB-I00, PID2023-147458NB-C21 and CEX2023-001292-S; by Generalitat Valenciana under contracts PROMETEO/2020/023 and  CIPROM/2023/59.

%\newpage
\bibliography{refs.bib} 
\end{document}